\documentclass[prd,showpacs,preprintnumbers,amsmath,amssymb,twocolumn,superscriptaddress,bibnotes
]{revtex4-1}
\usepackage{graphicx}
\usepackage{pbox}
\usepackage[normalem]{ulem}
\usepackage{color}
\usepackage[colorlinks=true,linkcolor=blue,citecolor=blue,linktoc=page,urlcolor=blue]{hyperref}

\newcommand\blue[1]{{\color{black}#1}}
\newcommand\red[1]{{\color{red}#1}}

\definecolor{orange}{RGB}{255,127,0}
\newcommand\orange[1]{{\color{orange}#1}}

\begin{document}

\title{Evidence of nodal superconductivity in LaFeSiH}

\author{A. Bhattacharyya}
\address{Department of Physics, Ramakrishna Mission Vivekananda Educational and Research Institute, Belur Math, Howrah 711202, West Bengal, India}
\author{P. Rodi\`{e}re}
\affiliation{Institut N\'eel, CNRS \& Univ. Grenoble Alpes, 38042 Grenoble, France}
\author{J.-B. Vaney}
\affiliation{ICMCB, CNRS \& Université Bordeaux, UPR 9048, F-33600 Pessac, France}
\author{P. K. Biswas}
\affiliation{ISIS Neutron and Muon Facility, SCFT Rutherford Appleton Laboratory, Chilton, Didcot Oxon, OX11 0QX, United Kingdom} 
\author{A. D. Hillier}
\affiliation{ISIS Neutron and Muon Facility, SCFT Rutherford Appleton Laboratory, Chilton, Didcot Oxon, OX11 0QX, United Kingdom}
\author{A. Bosin}
\affiliation{Dipartimento di Fisica, Universit\`a di Cagliari, IT-09042 Monserrato, Italy}
\author{F. Bernardini}
\affiliation{Dipartimento di Fisica, Universit\`a di Cagliari, IT-09042 Monserrato, Italy}
\author{S. Tenc\'{e}}
\affiliation{ICMCB, CNRS \& Université Bordeaux, UPR 9048, F-33600 Pessac, France}
\author{D. T. Adroja}
\email{devashibhai.adroja@stfc.ac.uk}
\affiliation{ISIS Neutron and Muon Facility, SCFT Rutherford Appleton Laboratory, Chilton, Didcot Oxon, OX11 0QX, United Kingdom}
\affiliation{Highly Correlated Matter Research Group, Physics Department, University of Johannesburg, PO Box 524, Auckland Park 2006, South Africa}
\author{A. Cano}
\email{andres.cano@cnrs.fr}
\affiliation{Institut N\'eel, CNRS \& Univ. Grenoble Alpes, 38042 Grenoble, France}

\date{\today}

\begin{abstract}

Unconventional superconductivity has recently been discovered in the first iron-based superconducting silicide LaFeSiH.
By using the complementary techniques of muon spin rotation, tunneling diode oscillator and density functional theory, we investigate the magnetic penetration depth and thereby the superconducting gap of this novel high-temperature superconductor. We find that the magnetic penetration depth displays a sub-$T^2$ behavior in the low-temperature regime below $T_c/3$, which evidences a nodal structure of the gap (or a gap with very deep minima). Even if the topology of the computed Fermi surface is compatible with the $s_\pm$-wave case with accidental nodes, its nesting and orbital-content features may eventually result in a $d$-wave state, more unusual for high-temperature superconductors of this class.
\end{abstract}

\maketitle

\section{Introduction}

The superconducting energy gap is a hallmark of superconductivity at the level of the electronic structure \cite{tinkham2004}. Furthermore, the symmetry of the gap function is intimately linked to the microscopic interactions that yield the Cooper pairing, 
thus providing key information about the mechanism behind superconductivity. Iron-based superconductors have proven to be a distinct class of unconventional superconductors \cite{hosono15} 
in which the gap symmetry can be tuned by means of external control parameters such as doping, pressure, or disorder \cite{hirschfeld11,hirschfeld2016}. In view of their distinct multiband features, it was soon realized that the so-called $s_\pm$-wave gap with a sign change between electron and hole pockets in the Fermi surface is the natural candidate for the gap function in most of these materials \cite{mazin-prl08}. In this case, doping, for example, can lead to enhanced anisotropy by means of various effects such as the modification of intraband Coulomb interactions and changes in the orbital weights on the Fermi surface. At the same time, it was also realized that the $d$-wave pair channel is a strong competitor to the $s_\pm$-wave one \cite{aoki08}. In this case the key role is played by the hole pockets where the gap function displays symmetry-imposed nodes. In fact, a strong tendency towards $d$-wave pairing, even dominating over the $s$-wave one, has been found in various models suited for 1111 systems, especially towards the overdoped limit \cite{aoki08,graser09,bernevig-09,bernevig-11,chubukov11}. 
These considerations explain the general trends observed in the Fe-based superconductors, including the controlled changes reported experimentally in BaFe$_2$(As,P)$_2$ \cite{Mizukami14} and (Ba,Rb)Fe$_2$As$_2$ \cite{Guguchia15}. 

Here, we investigate the gap structure of the novel superconductor LaFeSiH with $T_c \sim 10$~K in its parent phase \cite{bernardini-prb08}. This system is the first silicide in the family of Fe-based superconductors, whose unconventional mechanism of superconductivity is yet to be elucidated \cite{hung-prb18}. Compared to LaFeAsO, the shape of the Fermi surface is essentially preserved in LaFeSiH although it features an increased 3D character that reduces considerably the nesting (see \cite{bernardini-prb08} and Fig. \ref{Fig4} below). 
To determine the properties of the corresponding superconducting gap we measured the magnetic penetration depth $\lambda$. The temperature dependence of this fundamental quantity 
maps the excited quasiparticles, and hence the structure of the superconducting gap. 
Specifically, we performed muon spin rotation and relaxation ($\mu$SR) experiments and used tunnel diode oscillators (TDO) to determine $\lambda$. 
While the $\mu$SR technique provides a direct access to $\lambda$ by probing the magnetic field distribution in the vortex state \blue{(i.e. above $H_{c1}$)} \cite{sonier-rmp00, Bhattacharyya-18}, the TDO method enables the collection of a large density of points with very high resolution, and hence a very precise determination of the changes in $\lambda$ in the Meissner state \blue{(below $H_{c1}$)} \cite{Prozorov_2011}. These complementary techniques are supplemented with density-functional-theory (DFT) calculations, from which we compute the zero-temperature penetration depth $\lambda(0)$ in the London approximation and rationalize the nodal behavior observed in our measurements as a function of temperature.  

\blue{

\section{Methods \label{Methods}}

\subsection{Sample preparation} 

The LaFeSiH powder sample for the TF-$\mu$SR experiment was obtained as described in \cite{bernardini-prb08}. From this powder, small single-crystals were singled out for the TDO measurements.
The selected crystals have a slab geometry with typical thicknesses $2d\sim 10$~$\mu m$ in the $c$ direction and planar dimensions up to $2w\sim 300$~$\mu m$. 

\subsection{$\mu$SR experiment \label{A:muSR}}

The $\mu$SR experiment was carried out using the MuSR spectrometer at ISIS Facility, UK. Thus, we measured the muon spin depolarization that results from the application of a magnetic field of 30~mT ($>H_{c1}$, see \cite{bernardini-prb08}) in the transverse-field configuration (TF-$\mu$SR). This depolarization rate has a component due to the nuclear magnetic contributions of the sample. In addition, if the sample is a type-II superconductor, such a depolarization rate is expected to develop an extra contribution due to the inhomogeneous distribution of magnetic field in the vortex state, which is directly linked to the magnetic penetration depth $\lambda$ \cite{sonier-rmp00}.

\subsection{TDO measurements}

We used a high stability LC oscillator with resonant frequency 13~MHz driven by a tunnel diode in a $^3$He refrigerator. 
Thus, we measured the relative shift of the resonant frequency $\Delta f/\Delta f_0$ which is directly related to the AC magnetic susceptibility $\chi'$ and hence $\Delta \lambda (T) \equiv \lambda (T) - \lambda(0)$  (here $\Delta f_0$ is the frequency shift obtained when the sample is completely extracted from the coil at the base temperature, while the factor of proportionality is defined by the TDO effective dimension of the sample) \cite{carrington03,Prozorov_2011,Manzano02,Diener09}.

According to the size of the meaured samples, the TDO effective sample dimension is expected to be $\sim 0.2 w$ when the magnetic field is applied along the $c$ axis and $\sim d$ when it is perpendicular to $c$ \cite{Prozorov_2006}.
Furthermore, if $H\parallel c$ then the screening supercurrents flow entirely in the $ab$-plane and hence the in-plane penetration depth $\lambda_{ab}$ is probed. 
However, if $H\perp c$ the screening is due to supercurrents flowing both in-plane and out-of-plane so that the mixture $\lambda_{ab}+{ d\over w}\lambda_{c}$ containing the contribution due to the out-of-plane penetration depth $\lambda_{c}$ is probed.

\subsection{DFT calculations \label{A:DFT}}

We performed DFT calculations using the FLAPW method as implemented in the {\sc{WIEN2K}} package \cite{Wien2k} with the PBE exchange-correlation functional \cite{PBE}.
Specifically, we considered the low-temperature structure reported of LaFeSiH in \cite{bernardini-prb08}, with muffin-tin radii of 2.30, 2.10, 2.20 and 1.20 a.u for La, Fe,Si and H atoms respectively, and a plane-wave cutoff $R_{\rm MT}K_{\rm max}=5.0$ in our spinless calculations. 
The integration over the Brillouin zone was performed using a 15$\times$15$\times$7 $k$-mesh, 
while 
\blue{the Fermi surface was computed using a denser 64$\times$64$\times$32 $k$-mesh (the Fermi energy was determined by the tetrahedron method~\cite{Blochl}).
The Fermi velocity, in its turn, was computed on the dense mesh} 
as {\bf v} = {\bf p}$/m_e$, with {\bf p} being the expectation value of the momentum operator and $m_e$ the electron mass. 

From these calculations we further computed the penetration depth in the London approximation according to the formula 
$    
(\lambda_{ij}^{2})^{-1}(0)= \frac{\mu_0e^2}{4\pi^3\hbar}\oint_\text{FS} dS\frac{{ v}_i{ v}_j}{|{\bf v}|}
$
(see e.g. \cite{Prozorov_2006,Prozorov_2011}). Here $\bf v$ is the Fermi velocity and the integral is over the Fermi surface. In these calculations, we also computed the conductivity in the relaxation-time approximation which reads
$ \sigma_{ij}
= (e^2 \tau/\Omega_0)\int_{BZ} v_i({\bf k}) v_j({\bf k}) \delta(\varepsilon({\bf k}) - \varepsilon_F)  d{\bf k}
$ 
within the Boltzmann transport theory \cite{allen-88}.
Here $\Omega_0$ is the volume of the first Brillouin zone and the relaxation time $\tau$ gives the mean-free path as $\ell = v_{\rm F} \tau$.

}

\section{Results}

\subsection{$\mu$SR experiment}

\begin{figure*}[tbh!]
\includegraphics[height=0.24\textwidth]{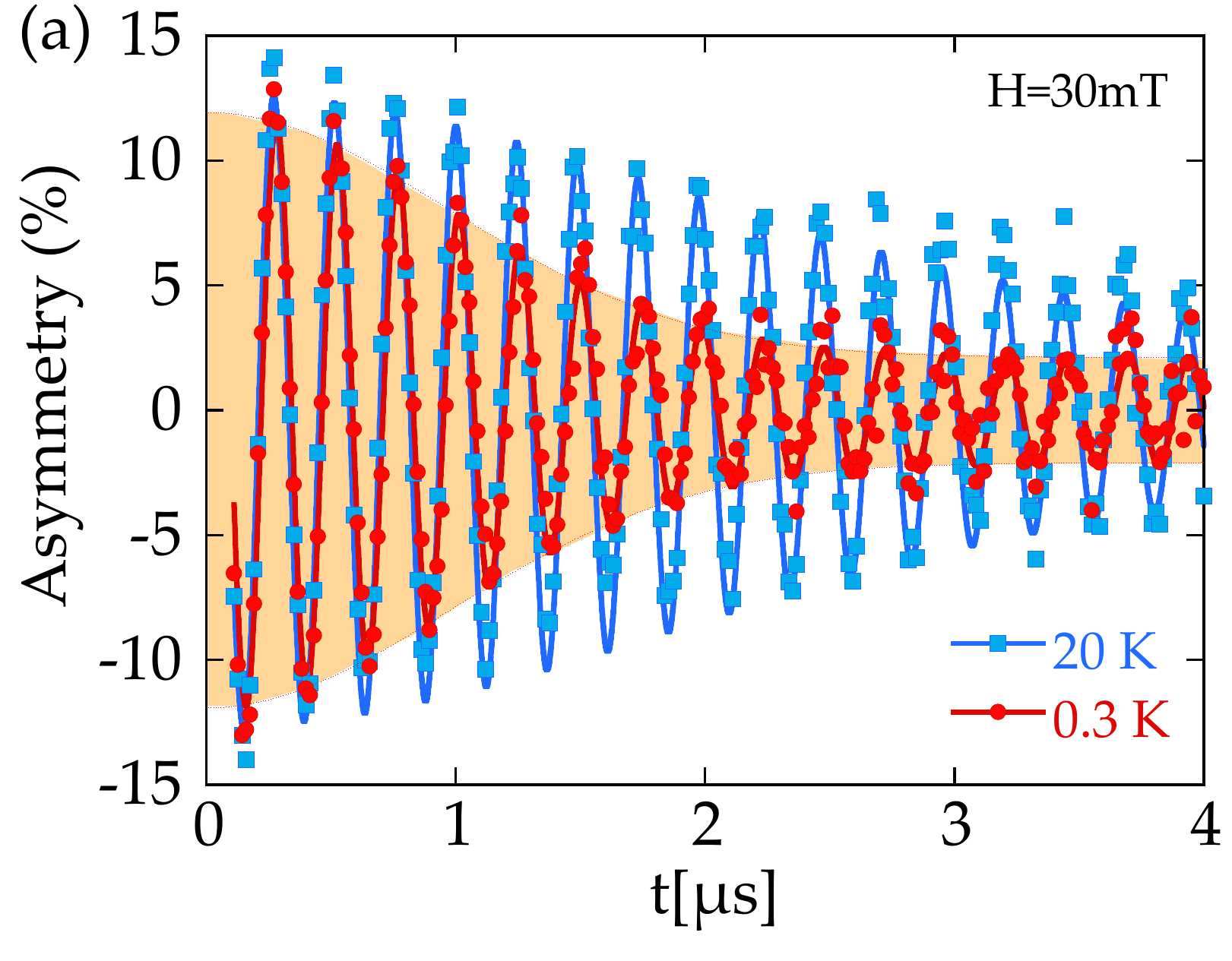}
\hspace{0.01\textwidth}
\includegraphics[height=0.24\textwidth]{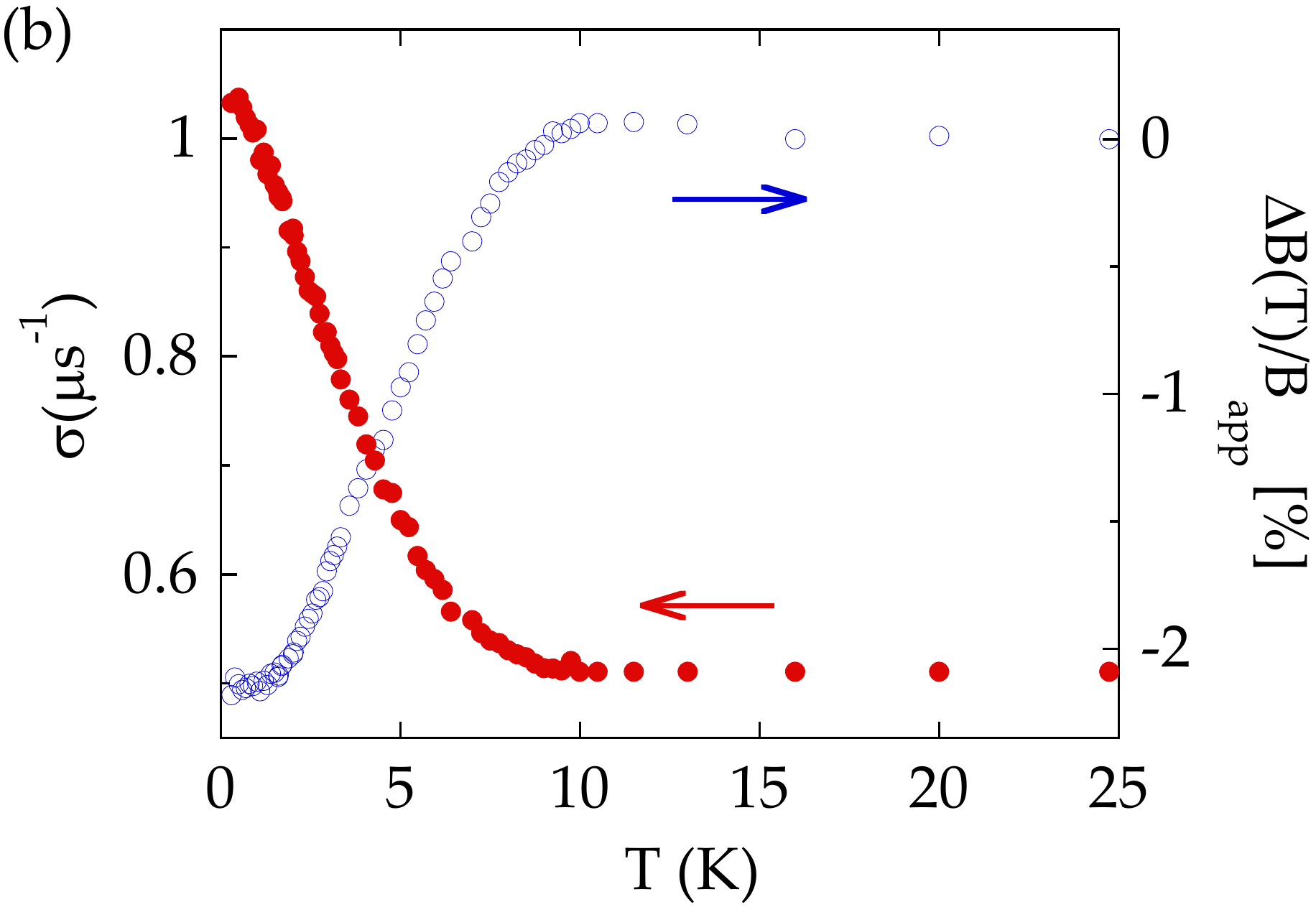}
\hspace{0.01\textwidth}
\includegraphics[height=0.24\textwidth]{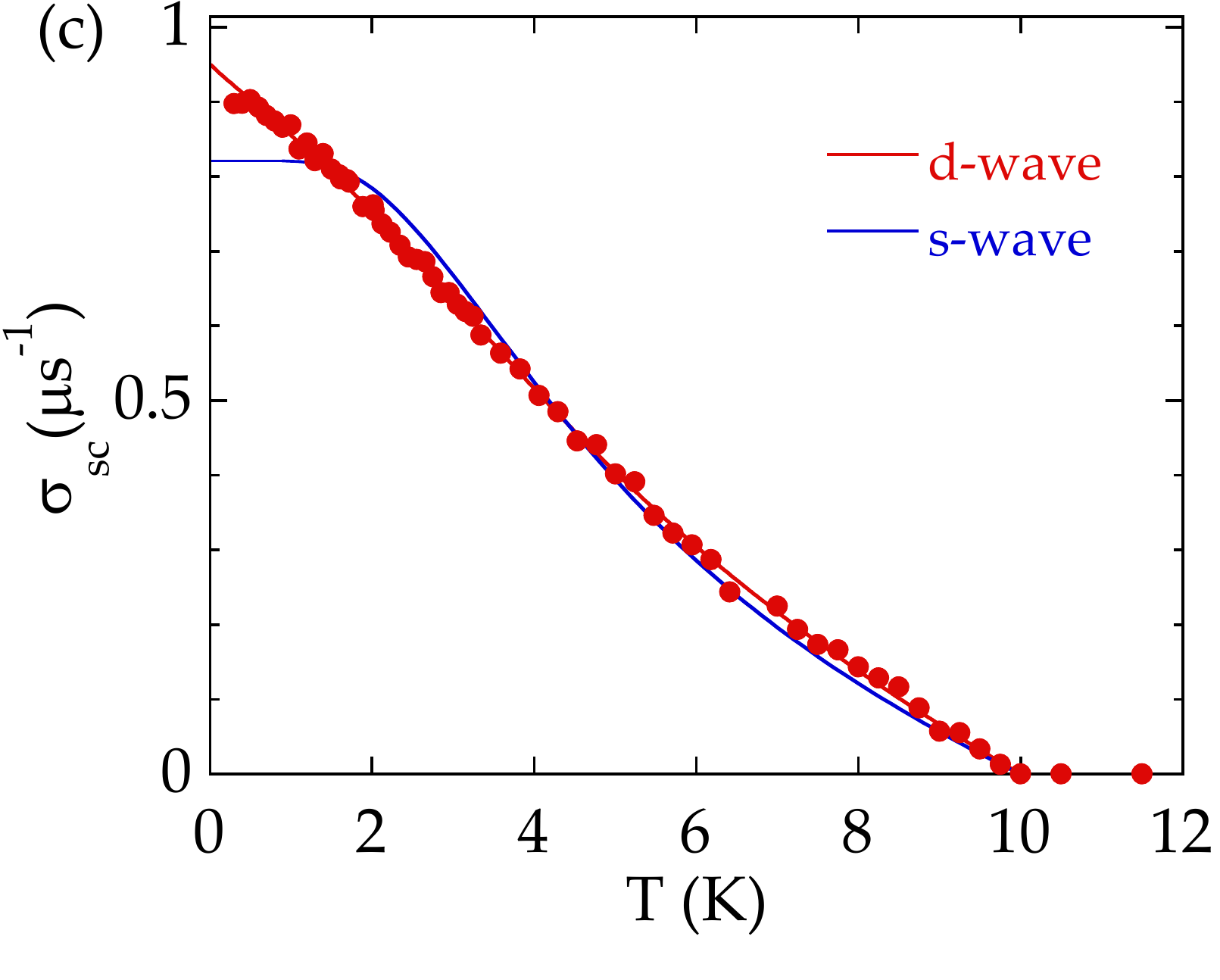}
\caption{
(a) TF-$\mu$SR  asymmetry spectra for LaFeSiH collected at $T$ = 20 K (blue) and at $T$ = 0.3~K (red) at an applied magnetic field of $H$ = 30 mT. 
The solid lines show the fits using Eq.~\eqref{MuonFit1} and the shaded area the envelop of the 0.3~K data. 
\blue{(b) Total muon depolarization rate  $\sigma$ and normalised internal field $\Delta B (T) /B_{\rm app} = [B(T)-B(T=20~{\rm K})]/B_{\rm app}$ with $B(T)=\omega_{s}(T)/\gamma_{\mu}$ as a function of temperature down to 0.3 K. The increase (decrease) of these quantities below 10~K reveals the emergence of superconductivity.
(c) Superconducting contribution $\sigma_{sc}(T)$ to the total muon depolarization rate. The lines illustrate the fits to the standard expressions that apply in the clean limit as discussed in the main text.} Even if some models can fit the data reasonably well, the resulting values of the gap are systematically below the BCS weak coupling limit thus indicating a convolution with the effect of impurities.}
\label{musr}
\end{figure*}

Firstly, we report the $\mu$SR experiment. Fig.~\ref{musr} (a) shows the transverse-field $\mu$SR (TF-$\mu$SR) asymmetry spectra measured
\blue{in powder LaFeSiH at $20$~K in the normal state and at $0.3$~K in the superconducting state}. The damping of the muon-time asymmetry oscillations observed in the normal state is very small, which indicates that these oscillations are mainly due to nuclear contributions with a distribution of the internal field that is extremely uniform at the applied field. In the superconducting state, in contrast, the damping is substantially higher, as expected from inhomogeneous field distribution created by the superconducting vortices.

\blue{We followed \cite{sonier-rmp00,Bhattacharyya-18} and modeled the TF-$\mu$SR asymmetry spectrum as 
\begin{equation}
A(t) = A_{s}
e^{-\sigma^{2}t^{2}/2}\cos(\omega_{s}t+\theta)
+
A_{bg}\cos(\omega_{bg}t+\theta).
\label{MuonFit1}
\end{equation}
The first term in this expression describes the oscillations (with relaxation) produced by the sample while the second accounts for the background oscillations (without relaxation) due to e.g. the Ag-sample holder, with $\theta$ being a phase related to the detector geometry. In the first term, the total relaxation rate $\sigma$ reads $\sigma = \sqrt{\sigma_{\rm nm}^2 +\sigma_{\rm sc}^2}$ where $\sigma_{\rm nm}$ and  $\sigma_{\rm sc}$ represent the aforementioned nuclear and superconducting vortex-lattice contributions respectively. Furthermore, for a triangular vortex-lattice in a type-II superconductor such that $\kappa=\lambda /\xi \gg70$ and  $0.13/\kappa^{2}\ll H/H_{c2}\ll1$, the superconducting part 
reduces to $\sigma_{\rm sc}^2=3.71\times 10^{-3} \frac{\gamma_\mu^2\phi_0^2}{\lambda^4}$, where $\gamma_\mu$ is the muon gyromagnetic ratio and $\phi_0$ is the flux quantum 
~\cite{sonier-rmp00,amato,Brandt}.
The normalised internal field, in its turn, can be estimated from the $\mu$SR data as $\Delta B (T) /B_{\rm app} = [B(T)-B(T_n=20~{\rm K})]/B_{\rm app}$ with $B(T)= \omega_{s}(T)/\gamma_{\mu}$.}

The lines in Fig.~\ref{musr}(a)
illustrate the fits of the TF-$\mu$SR data according to Eq. \eqref{MuonFit1}. 
The parameters $A_{bg}$ and $\theta$ were estimated by fitting to the 0.3 K data and 20 K respectively and their values were kept fixed in the fitting of the other temperature data points (the background amplitude was 18\% of the total one). The parameters $A_{s}$ and $\omega_{bg}$ were allowed to vary, which nevertheless resulted in nearly temperature-independent parameters. The good agreement of the fits validates the model Eq. \eqref{MuonFit1}. 

The total depolarization rate \blue{and the normalised internal fied are plotted in Fig.~\ref{musr}(b) as a function of temperature. 
These quantities are almost constant above $T_c$. The nuclear contribution to $\sigma$, in particular, can then be estimated as $\sigma_{\rm nm}=0.508(1)$ \blue{$\mu$s$^{-1}$}. The relatively large value of this relaxation rate can be due to additional muon-H interactions (see e.g. \cite{kadono08}) or the presence of Fe impurities. Beyond that, the clear changes observed in both $\sigma$ and $\Delta B$ with decreasing the temperature confirms the emergence of superconductivity in LaFeSiH. 
The estimated superconducting contribution $\sigma_{\rm sc}$ is shown in Fig.~\ref{musr}(c). 
}
The $T_c$ derived from this data is $ \simeq 10$~K, which is slightly higher than the onset observed in the DC magnetic-response measurements on the same samples (see also \cite{bernardini-prb08}). Since $\mu$SR has much higher sensitivity with respect to the superconducting volume fraction, this suggests that there is a non-negligible distribution of $T_c$'s within the sample.

\subsubsection{Zero-temperature penetration depth $\lambda(0)$}

The zero-temperature penetration depth $\lambda(0)$ can be directly determined from the TF-$\mu$SR parameter $\sigma_{\rm sc}$. The extrapolation of $\sigma_{\rm sc}$ to zero temperature gives $\lambda(0) = 336$~nm, which is similar to that reported in other Fe-based superconductors \blue{and correlates well with a $T_c$ of 10~K as expected from the Uemura-plot phenomenology of high-temperature superconductors \cite{uemura-prl89,luetkens-prl08,bernardini20}}. This quantity, however, has to be understood as the effective penetration depth $\lambda_\text{eff} \approx 3^{1/4}[1 +2(\lambda_{ab}/\lambda_{c})]^{-1/4}\lambda_{ab}$ \cite{fesenko1991}. \blue{In anisotropic layered compounds this quantity is generally dominated by the in-plane penetration depth $\lambda_{ab}$, so that $\lambda_\text{eff} \approx 3^{1/4}\lambda_{ab}$. In our case, that would mean that $\lambda_{ab}\approx255$~nm. However, as we show in Sec. \ref{sec:discussion} below, the actual anisotropy is comparatively moderate in LaFeSiH and, more importantly, the theoretical value of the $\lambda_\text{eff}$ is noticeably smaller than the one deduced from the $\mu$SR data.
The latter}
difference could be ascribed to the scattering to impurities \cite{tinkham2004}. 

\subsubsection{Temperature dependence of $\lambda$}

When trying to fit the overall temperature dependence of $\sigma_{\rm sc}\propto \lambda^{-2}$ to the standard expression that would be applicable in the clean limit, we find that the fits yield gap values that are systematically below the BCS weak-coupling limit. 
\blue{Specifically, we tried to fit the data according to 
$\frac{\sigma_{sc}(T)}{\sigma_{sc}(0)} =  1 + \frac{1}{\pi}\int_{0}^{2\pi}\int_{\Delta(T,\phi)}^{\infty}\Big(\frac{\partial f}{\partial E}\Big) \frac{EdEd\phi}{\sqrt{E^{2}-\Delta(T,\phi)^2}}$.
Here $f = [1+{\rm exp}(E/k_B T)]^{-1}$ is the Fermi function, $\phi$ is the azimuthal angle across the Fermi surface, and $\Delta(T,\phi)$ is the superconducting gap function. As customary, we expressed the latter as $\Delta(T,\phi) = \Delta_{0}\Gamma(T/T_c)g(\phi)$, with  $\Gamma(T/T_c)$ = tanh$\{1.82[1.018(T_c/T-1)]^{0.51}\}$ and 
$g(\phi)=1$ for describing an isotropic $s$-wave gap and $g(\phi)=|\cos(2\phi)|$ for a $d$-wave gap with line nodes \cite{carrington03,Bhattacharyya-18}. We also considered other gap structures as well as multi-gap extensions of this model. 
The $d$-wave model provides the best fit among the one-gap models but implies an unphysical gap $\Delta_0 = 1.41 k_BT_c < 1.764 k_BT_c$ ($\Delta_0 = 0.84 k_BT_c $ in the $s$-wave case). Similarly, $s+s$- or $s+d$-wave models imply unphysical gaps $(\Delta_{0s}, \Delta_{0s}) = (1.04, 0.36) k_BT_c$ and $(\Delta_{0s}, \Delta_{0d}) = (1.09, 0.66) k_BT_c$ below the BCS limit all of them.
}

This \blue{exercise} confirms that the scattering with impurities \blue{does} play a role and convolutes with such a nominal temperature dependence \blue{as described in} \cite{Prozorov_2011}. Thus, we focus on the low-temperature behavior. Fig. \ref{Fig2} shows the changes in $\lambda(T)$ as a function of $T^2$ in the low-temperature limit ($T<T_c/3$). The $\mu$SR data readily suggests a sub-$T^2$ behavior, and thereby the presence of line nodes in the superconducting energy gap (see e.g.  \cite{Prozorov_2011,hirschfeld2016} and the discussion below). 

\subsection{TDO measurements}

In order to confirm the temperature dependence of $\lambda$ revealed by the $\mu$SR experiment we performed additional TDO measurements on single-crystals. This enables, in particular, the collection of a much higher density of data points. Fig. \ref{Fig3} shows the AC magnetic susceptibility measured in LaFeSiH as a function of the temperature when the magnetic field applied along the $c$-axis and in the basal $ab$-plane. The perfect diamagnetic behavior observed in both cases confirms the superconducting transition with onset $T_c \approx 10$~K. \blue{
Above $T_c$ in the normal state, the TDO signal becomes constant and the {\it in-situ} extraction of the sample reveals that the AC magnetic field is not screened. 
Induced eddy currents are expected to be distributed within the sample with a skin depth $\delta=\sqrt{\rho/(\pi\mu_0 f)}$, where $\rho$ is the resistivity, $\mu_0$ the magnetic permeability, and $f$ the AC frequency (13~MHz in our case). In our measurements we observed no detectable variation of the resonant frequency above $T_c$ ---as could be the case if $\delta $ changes due to changes in $\rho$. Thus, we conclude that $\delta$ is always larger than the size of the sample and, accordingly, we estimate} the normal-state resistivity as $>$ 20 $\mu\Omega$~cm.

\begin{figure}[t!]
\includegraphics[width= 0.4\textwidth]{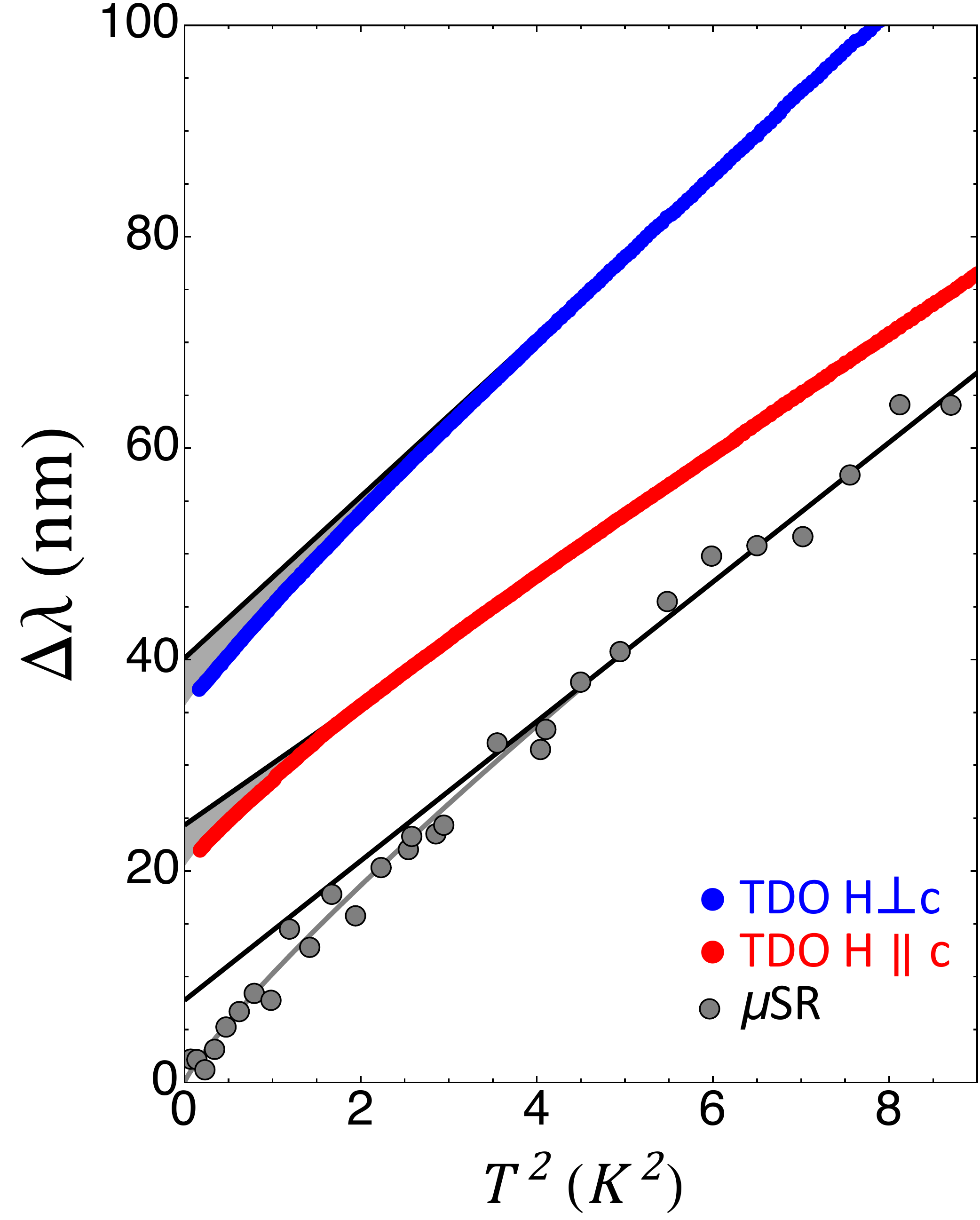}
\caption{
Change in the magnetic penetration depth $\Delta \lambda$ of LaFeSiH against $T^2$ in the low-temperature region below $T_c/3$. The $\mu$SR data (powder) is shown by the gray circles while the TDO data (single-crystal) is in red for $H\parallel c$ and in blue for $H\perp c$. The TDO data has been normalized according to the $\mu$SR data and a vertical offset has been introduced for clarity. Black lines are $T^2$ fits above 2.5~K. The two data sets clearly follow a power-law $T^n$ behavior with $n < 2$ at low temperatures, revealing the presence of line nodes in the superconducting gap. } 
 \label{Fig2}
\end{figure}

\begin{figure}[t!]
\includegraphics[width= 0.4\textwidth]{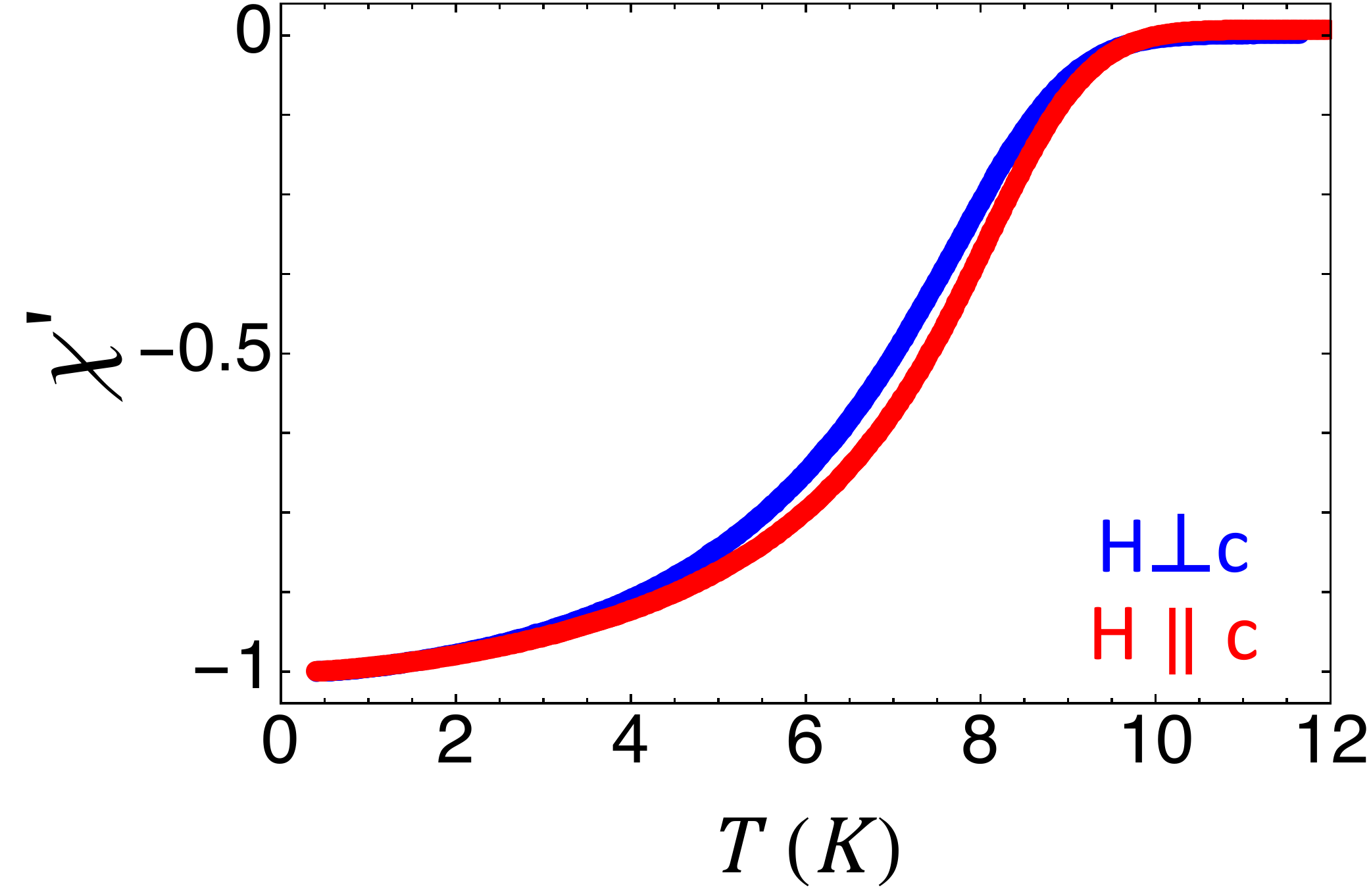}
\caption{Magnetic response measured in single-crystal LaFeSiH. AC susceptibility across the superconducting transition ($T_c \approx 10$~K) with the magnetic field applied along the $c$-axis (red) and in the basal $ab$-plane (blue). } 
 \label{Fig3}
\end{figure}

The magnetic susceptibility displays essentially the same behavior as a function of the temperature irrespective of the direction of the applied field. This suggests that the anisotropy between in-plane and out-of-plane superconducting properties is such that $ \Delta \lambda_{ab} > \Delta\lambda_{c}/30 $, so that the signal is always dominated by $ \Delta \lambda_{ab}$ due to the aspect ratio of the samples ($d/w \sim 1/30$). This is in fact in tune with the weak anisotropy obtained in our DFT calculations (see Table \ref{LondonDFT}). At the same time, the quantitative agreement between the susceptibility for the two orientations of the magnetic field is rather surprising. Also, the drop across the transition is quite broad indicating again a non-negligible distribution of $T_c$'s. These features suggest that the effective geometrical factors are more complex in these samples. 

Fig. \ref{Fig2} shows the measured changes in the magnetic penetration depth as a function of $T^2$ in the low-temperature limit (TDO data is in red and blue). These changes confirm the sub-$T^2$ behavior observed in the $\mu$SR data.  
Specifically, when the data is fitted over $T < T_c/3 =3$~K, the exponent is found $n =1.8$ for the magnetic field applied along $c$ and $1.7$ for the field in the perpendicular direction. These values do not change when the fitting interval is reduced to $T<T_c/6 =1.5$~K for example, and the same sub-$T^2$ behaviour is observed in other equivalent samples.

\section{Discussion \label{sec:discussion}}

The linear-in-$T$ behavior at $T\ll T_c$ of the penetration depth of a superconductor with line nodes is well known to become $T^2$ due to impurity scattering \cite{Prozorov_2011,hirschfeld2016}. In the case of a fully gapped superconductor with an unconventional gap structure such as the $s_\pm$ one, the exponential behavior also becomes $T^2$ due to impurities. However, the latter possibility is ruled out in our case since that would yield an exponent $n \geq 2$ while we always observe $n\leq 2$ in our experiments (both $\mu$SR and TDO). 
Likewise, an extended $s$-wave 
with $c$-axis line nodes can be ruled out. The sub-quadratic behaviour, however, is compatible with either a $s_\pm$-wave with more general accidental nodes ---or very deep gap minima--- or a $d$-wave with symmetry-imposed nodes, both in the presence of impurities.

Regarding the nature of these impurities, we note that the measured $\lambda$ does not display any Curie upturn at low temperatures. This indicates that they are non-magnetic. The degree of disorder introduced by these impurities can be quantified by comparing the zero-temperature coherence length $\xi_0 \simeq 4.3$~nm \cite{bernardini-prb08} to the mean-free path $\ell$. The latter quantity can be estimated from the measured value of the normal-state conductivity and the one computed from DFT as described in \ref{A:DFT}, which correctly captures the complex multiband features of our system.
Thus, $\ell$ is estimated to be $\lesssim 5$~nm. According to this estimate, the samples seem to be in a borderline case between the clean ($\ell \gg \xi_0$) and the dirty limit ($\ell \ll \xi_0$). This obviously makes the quantitative analysis of the data rather involved, which can explain the aforementioned limitations related to the $\mu$SR fits and the TDO geometrical factors. In any case, both these data sets display the same sub-$T^2$ behavior revealing the nodal character of the superconducting gap in LaFeSiH.

This can be further discussed in relation to the corresponding Fermi surface. The 1111 compound LaFeAsO provides a reference electronic structure for the Fe-based superconductors. Here, the Fermi surface displays electron and hole pockets that are separated by the wavevector $(\pi, 0)$ (in the 1Fe/unit-cell notation), so that standard considerations on pairing by repulsive interactions suggest a $s_\pm$-wave pair state \cite{mazin-prl08} with a subdominant $d$-wave channel \cite{aoki-prl09,graser09,bernevig-09,hackl-18}. In this picture the anisotropy in the $s_\pm$ gap function is controlled by several features, notably the 2D vs. 3D character of the Fermi surface and its orbital weights \cite{hirschfeld-prb09a,schmalian-prb09}. 
The strength of the $(\pi,0)$ interactions, in particular, has a strong dependence on the presence/absence of a $d_{x^2-y^2}$ band near the Fermi level (eventually determined by the actual lattice structure), which then controls the nodeless vs. nodal character of the $s_\pm$ state and can even promote the $d$-wave one \cite{aoki-prb09}. 

\begin{figure}[t!]

{\footnotesize
\hspace{3em}\red{$d_{xy}$} \; \orange{$d_{xz/yz}$} \; \blue{$d_{3z^2-r^2}$} }

\includegraphics[height=.275\textwidth]{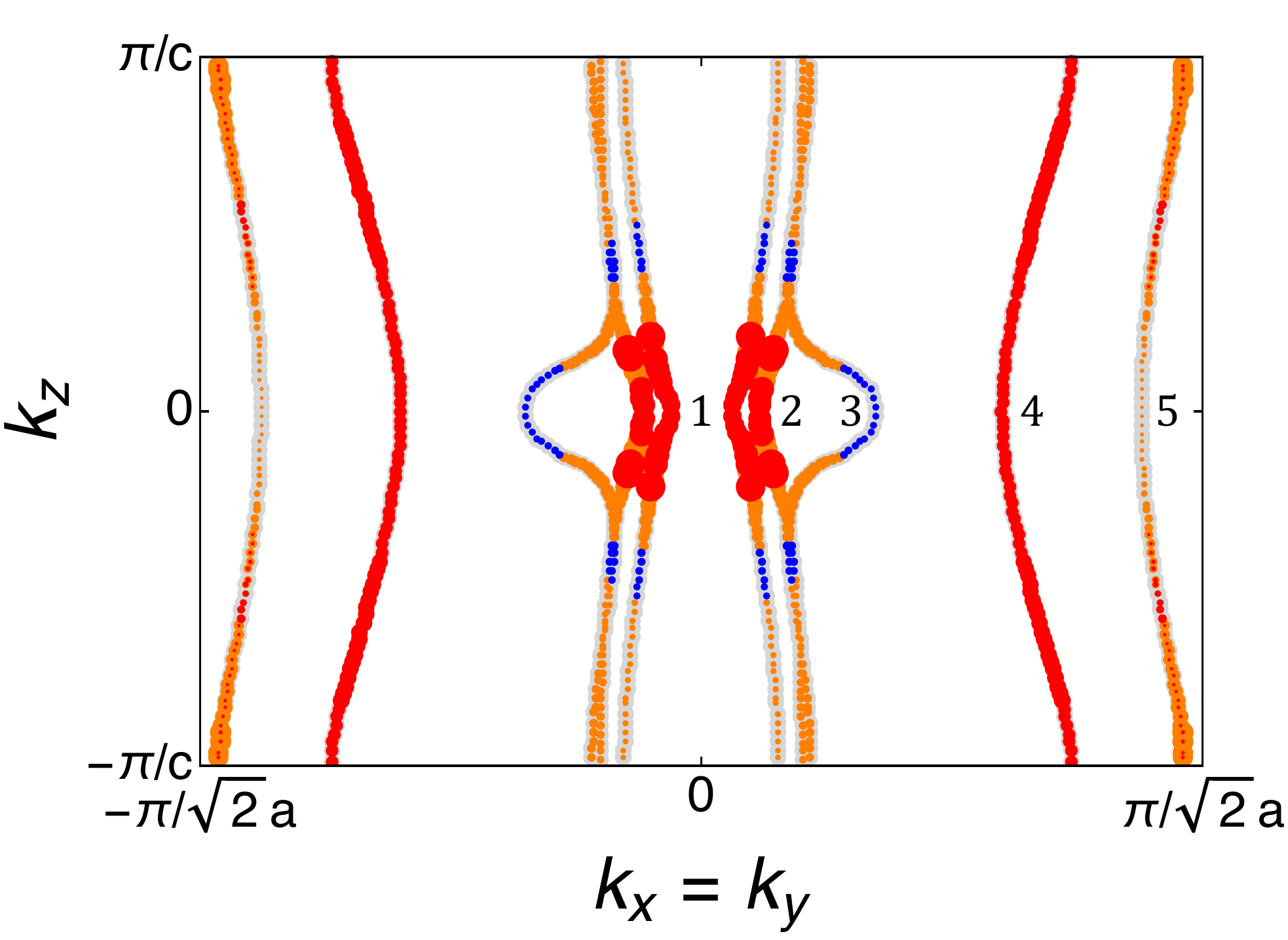}


\includegraphics[height=.275\textwidth]{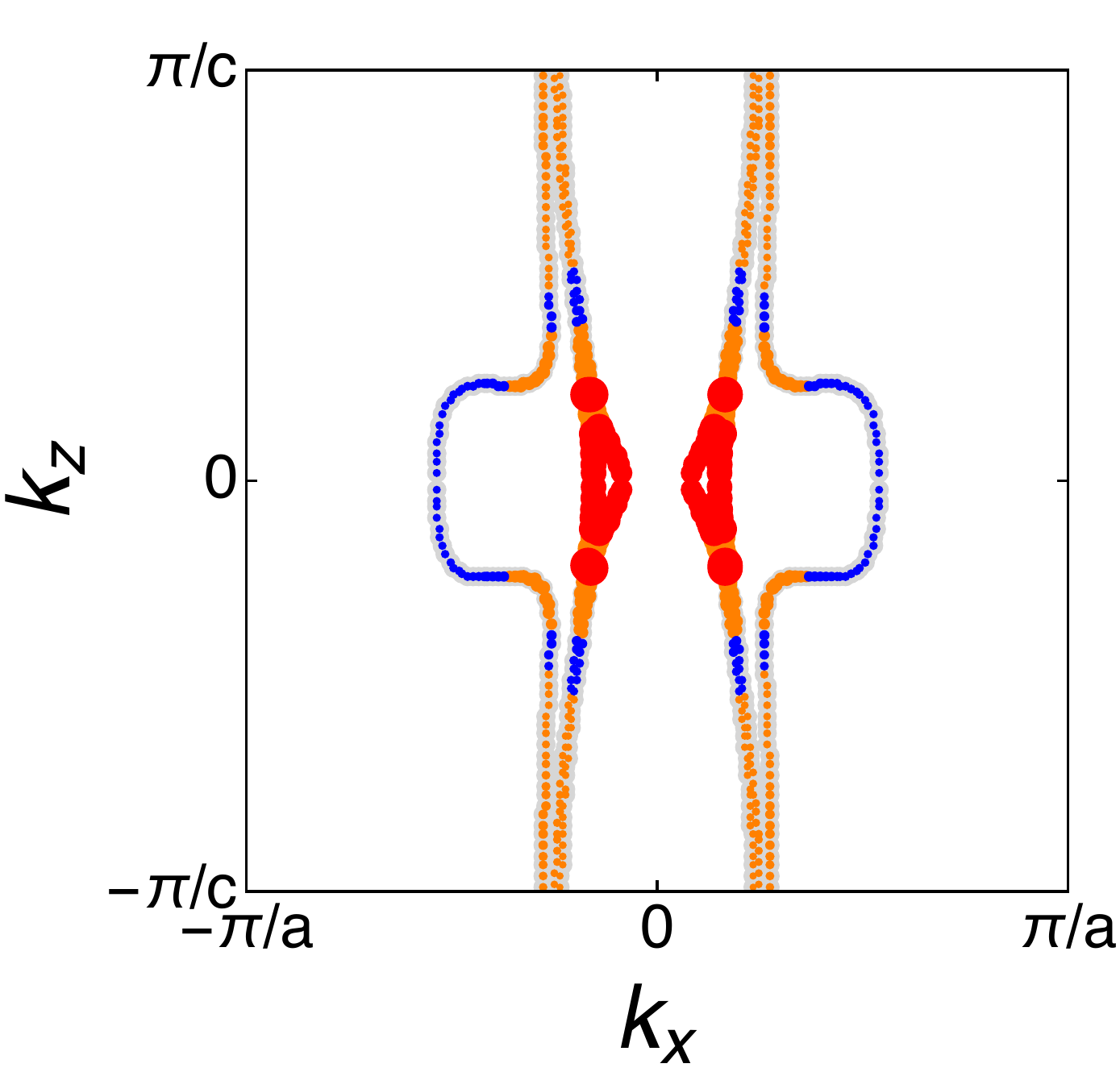}

 \caption{Fermi-surface cuts of LaFeSiH at $k_x=k_y$ (top) and $k_y=0$ (bottom). The size of the circles scales with the  contribution of the orbitals that are indicated by the different colors: $d_{xy}$ (red), $d_{xz/yz}$ (orange), and $d_{3z^2-r^2}$ (blue). The Fermi sheets are numbered in the top panel.} 
 \label{Fig4}
\end{figure}

\begin{table}[t!]
 \begin{tabular}{c c c c c c c}
\hline \hline    
FS sheet & 1 & 2 & 3 & 4 & 5 & Total\\ 
\hline $\lambda_{ab}(0)$ (nm) & 323 & 266 & 207 & 145 & 226 & 94 \\
$\lambda_c(0)$ (nm) & 698 & 951 & 186 & 305 & 759 & 150 \\ 
\hline \hline
\end{tabular}
\caption{Zero-temperature magnetic penetration length of LaFeSiH obtained from DFT calculations in the London approximation. The different columns indicate the values obtained from each Fermi-surface sheet, labelled as in Fig. \ref{Fig4} (sheets 1-3 and 4-5 correspond to the hole-like and electron-like pockets respectively in the $k_z=0$ plane, i.e. around $\Gamma$ and M). \blue{Taking into account that $\lambda^{-2}$ is proportional to the superfluid density  ---so that the additive quantity is $\lambda^{-2}$ rather than $\lambda$--- the total values are obtained as $\lambda_\text{tot} = (\sum_{n}\lambda_{n}^{-2})^{-1/2}$ (here $n=1, 2,\dots 5$ refers to Fermi-surface sheets). This further yields $\lambda _\text{eff} = 100$~nm.}
}
\label{LondonDFT}
\end{table}

Compared to LaFeAsO, the electronic band structure of LaFeSiH is visibly more 3D as illustrated by the Fermi surface cuts shown in Fig. \ref{Fig4}. 
This is largely due to the prominent \blue{$k_z$ dispersion of inner Fermi-surface sheets 3 and 1 (hole-like pockets) and also the outer one 4
(electron-like). This results into a weakened anisotropy in the calculated total $\lambda$ (see Table \ref{LondonDFT}). Beyond that,} 
the overall nesting between electron and hole pockets is drastically deteriorated, which is highly detrimental for the fully gapped $s_\pm$ pairing and can introduce accidental nodes \cite{aoki-prb09,bernevig-11b}. In addition, the $d_{x^2-y^2}$ character of the outer sheets is absent, which also goes in the same direction. The $d$-wave channel, in contrast, is mainly linked to the electron pockets (sheets 4 and 5) which better retain its propitious features. These considerations support our experimental finding of a nodal superconducting gap in LaFeSiH, including the comparatively rare $d$-wave case \cite{aoki08,carrington-09} among the Fe-based superconductors as a possible candidate.

\section{Conclusions} 
In conclusion, we have determined the magnetic penetration depth of the novel iron-based superconducting silicide LaFeSiH as a function of the temperature in the vortex and in the Meissner state using muon-spin rotation and tunnel-diode oscillators. The observed power-law behavior reveals the presence of low-energy exitations characteristic of nodal superconductivity. The effective zero-temperature value is found to be $\lambda(0)=336$~nm, which is consistent with \blue{the lower bound of} DFT calculations. 
The specific features of the electronic band structure of LaFeSiH suggest a prominent role of the electron pockets in the Cooper pairing and accordingly a $d$-wave superconducting state, even if a $s_\pm$-wave superconducting gap with accidental nodes (or more generally deep gap minima) is also compatible with our experimental findings. 
This outlines an analogy to the overdoped behavior of previous iron-based superconductors that is expected to motivate further studies.

\begin{acknowledgments}
P.R., F.B., J.B.V., S.T, and A.C. are supported by the Grant ANR-18-CE30-0018-03 IRONMAN. DTA and ADH acknowledge financial assistance from CMPC-STFC grant number CMPC-09108. AB would like to acknowledge the Department of Science and Technology (DST) India, for an Inspire Faculty Research Grant (DST/INSPIRE/04/2015/000169), and the UK-India Newton grant for funding support. F.B. acknowledges the Visiting Scientist Program
of the Centre de Physique Theorique de Grenoble-Alpes (CPTGA) for financial support. We would like to thank G. Stenning and D. Nye for their help in the sample characterization and the ISIS Facility for proving beam time on the MuSR spectrometer, DOI: 10.5286/ISIS.E.RB1900103.
\end{acknowledgments}

\bibliography{bib.bib}

\end{document}